\documentclass[twoside]{dis08}
\usepackage[latin1]{inputenc}
\usepackage[dvips]{graphicx,epsfig,color}
\usepackage{wrapfig,rotating}
\usepackage{amssymb,amsmath,array,cite}
\newcommand\GeV{\,\mbox{GeV}}
\newcommand\MeV{\,\mbox{MeV}}

\begin{document}
\title{{
\small \rm DESY 08-083 \hfill SFB-CPP-08-40 }\\
Higher Twist Contributions to \\ Deep-Inelastic 
Structure Functions}

%***********************************************************************
% AUTHORS INFORMATION AREA
%***********************************************************************
\author{Johannes Bl\"umlein$^1$ and Helmut B\"ottcher$^1$
%
% Optional short acknowledgment: remove next line if non-needed
\thanks{
This work was supported in part by by DFG Sonderforschungsbereich Transregio 9,
Computergest\"utzte Theoretische Physik.}
%
% DO NOT MODIFY THE FOLLOWING '\vspace' ARGUMENT
\vspace{.3cm}\\
%
% Addresses and institutions (remove "1- " in case of a single institution)
1- Deutsches Elektronen-Synchrotron, DESY \\
Platanenallee 6, D-15738 Zeuthen, Germany
%
% Remove the next three lines in case of a single institution
%\vspace{.1cm}\\
%2- School of Second Author - Dept of Second Author \\
%Address of Second Author's school - Country of Second Author's school\\
}
%***********************************************************************
% END OF AUTHORS INFORMATION AREA
%***********************************************************************

\maketitle

\begin{abstract}
We report on a recent extraction of the higher twist contributions to the deep
inelastic structure functions $F_2^{ep,ed}(x,Q^2)$ in the large $x$ region. It is 
shown that the size of the extracted higher twist contributions is strongly correlated 
with the higher order corrections applied to the leading twist part. A gradual lowering 
of the higher twist contributions going from NLO to N$^4$LO is observed, where in the 
latter case only the leading large $x$ terms were considered. 
\end{abstract}
%%%%%%%%%%%%%%%%%%%%%%%%%%%%%%%%%%%%%%%%%%%%%%%%%%%%%%%%%%%%%%%%%%%%%%%
%	Introduction
%%%%%%%%%%%%%%%%%%%%%%%%%%%%%%%%%%%%%%%%%%%%%%%%%%%%%%%%%%%%%%%%%%%%%%%
%\section{Introduction}
%\label{sec:introduction}
%%%%%%%%%%%%%%%%%%%%%%%%%%%%%%%%%%%%%%%%%%%%%%%%%%%%%%%%%%%%%%%%%%%%%%%

\vspace{1mm}\noindent
In wide kinematic regions the deeply inelastic structure functions can be 
described by their leading twist contributions within Quantum Chromodynamics (QCD). 
Higher twist corrections \cite{HT} emerge both in the region of large \cite{HT1a,HT1} 
and small values \cite{HT2,HT3} of the Bjorken variable $x$. The leading twist sector, 
both for 
unpolarized and polarized deeply inelastic scattering, is well explored within 
perturbative 
QCD up to the level of 3--loop, resp. 2--loop,  corrections \cite{THR1,THR2}, including the 
heavy flavor contributions \cite{HEAV}. On the other 
hand, very little is known on the scaling violations of dynamical next-to-leading twist 
correlation functions and the associated Wilson coefficients \cite{HT}, even on the 
leading order level. In many experimental and 
phenomenological analyzes, cf.~\cite{HT1a,HT1}, higher twist contributions are parameterized 
by an `Ansatz' \cite{HT1a}, which is fitted accordingly. Within QCD this ad-hoc treatment  
cannot  be justified, performing at the same time a higher order analysis for the 
leading twist terms. Since neither the corresponding higher twist anomalous dimensions nor
Wilson coefficients were calculated, the data analysis has to be limited in the first place 
to the kinematic domain in which higher twist terms can be safely disregarded. 

In the following we report on the determination of the higher twist contributions in the 
deeply-inelastic structure functions $F_2^{ep,ed}(x,Q^2)$, see Ref.~\cite{BBHT} for 
details. Higher twist contributions were also studied in deep-inelastic neutrino 
scattering, cf.~\cite{NUHT}. Also in the case of polarized deeply inelastic scattering
higher twist corrections are present in general. Since the polarized structure functions 
are measured through an asymmetry, the effect of higher twist 
contributions in the denominator function has to be known in detail. In \cite{BBpol} no 
significant higher twist contributions were found. Other authors claim contributions in 
the low $x$ region~\cite{LEAD}, which is also the region of very low values of $Q^2$.
These effects need further study.

In the case of the structure functions $F_2^{ep,ed}(x,Q^2)$ we investigate the
flavor non-singlet contributions in the large $x$ region for $Q^2 \geq 4 \GeV^2, W^2 
\geq 12.5 \GeV^2$, cf. \cite{BBG}.~\footnote{For 
power correction analyzes in the resonance region see \cite{RESO}. Here the concept of
the twist-expansion is not applicable, except assuming duality.} One generally may 
consider flavor non-singlet combinations and perform a three-loop QCD analysis,
which requires the $O(\alpha_s^2)$ Wilson coefficients \cite{ZN} and the 3--loop anomalous 
dimensions \cite{THR1}. The analysis can even be extended effectively to 4--loop order, since 
the dominant contribution there is implied by the 3--loop Wilson coefficient \cite{THR2}, 
parameterizing the yet unknown 4--loop anomalous dimension with a $\pm$ 100~\% error added to   
an estimate of this quantity formed as Pad\'e-approximation out of the lower order terms.
A comparison with the 2nd moment of the 4--loop anomalous dimension calculated in \cite{BC}
showed \cite{BBG} that the agreement is better than 20~\%, which underlines that the above 
approximation may be possible. We limit the QCD--analysis of the twist--2 contributions
to this representation since neither $\alpha_s(\mu^2)$ nor the splitting and coefficient functions 
are known beyond this level. Furthermore, heavy quark and target mass corrections 
are applied.

The evolution equations are solved in Mellin-$N$ space, cf.~\cite{BV}. The non--singlet 
structure function at the starting scale of the evolution, $Q^2_0$, is given by
%-----------------------------------------------------------------------
\begin{eqnarray}
\label{eq1}
F_2^{p,d; \rm NS}(N,Q^2) = \sum_{k=0}^\infty a_s^{k-1}(Q^2) C_{k-1}^{\rm NS}(N)
             f_2^{p,d; \rm NS}(N,Q^2)~,
\end{eqnarray}
%-----------------------------------------------------------------------
with $C_k^{\rm NS}(N)$ the expansion terms of the non--singlet Wilson coefficient with 
$C_0(N) = 1$, $a_s(Q^2)= \alpha_s(Q^2)/(4\pi)$ and $f_2^{p,d; NS}(N,Q^2)$ the corresponding
combination of quark distributions, cf. \cite{BBG}. Here we identify both the 
renormalization and
factorization scale with $Q^2$. Beyond $O(a_s^3)$ dominant large $x$ contributions to the
Wilson coefficient were calculated in \cite{resum}.

We then extrapolate the 
results to the region $4 \GeV^2 \leq W^2 \leq 12.5 \GeV^2$ and determine effective  higher twist coefficients
$C_{\rm HT}(x,Q^2)$ given by
%------------------------------------------------------------------------------------------------ 
\begin{equation}
F_2^{\rm exp}(x,Q^2) = F_2^{\rm tw2}(x,Q^2) \cdot \left[
\frac{O_{\rm TMC}\left[F_2^{\rm tw2}(x,Q^2)\right]} 
                      {F_2^{\rm tw2}(x,Q^2)} 
+ \frac{C_{\rm HT}(x,Q^2)}{Q^2 [1 \GeV^2]}\right]~.
\end{equation}
%------------------------------------------------------------------------------------------------ 
Here $O_{\rm TMC}[~~]$ denotes the operator of target mass corrections.

QCD corrections beyond N$^3$LO are known in form of the dominant large-$x$ contributions 
to the QCD--Wilson coefficients \cite{resum}. Since these corrections do quantitatively 
only apply in the range of large $x$ we do not 
use them in the twist-2 QCD--fit, because here the data are mainly situated at lower 
values of $x$ where beyond 4--loop order 
other contributions to the Wilson coefficients, which are not calculated yet, are as 
important. Furthermore, the 4--loop anomalous dimensions are yet unknown beyond the 
2nd moment. The leading large $x$ contributions are given in terms of harmonic sums of 
the type $S_{1,1, \ldots, 1}(N)$ which obey a determinant representation \cite{BK}
in single harmonic sums $S_l(N)$.
The effective higher twist distribution functions $C_{\rm HT}^{p,d}(x)$ extracted are shown in Figures~1  from
NLO to N$^4$LO.
Here we averaged over the values in $Q^2$ within the $x$--bins.
The leading twist terms are those given in \cite{BBG}, with the values of $\Lambda_{\rm QCD}^{(4)} = 
265 \pm 27, 226 \pm 25, 234 \pm 26 \MeV$, resp. in NLO, NNLO, and N$^3$LO.
Both for the proton and deuteron data $C_{\rm HT}(x)$ grows towards large values of $x$, 
and takes values $\sim 1$ around $x = 0.6$. The inclusion of higher order corrections reduces 
$C_{\rm HT}(x)$ to lower values with a gradually
smaller difference order by order. 
Yet for the highest bins, $x \geq 0.8$, the effect of the large $x$ resummation
terms is important. Earlier higher twist analyzes \cite{HT1} limited to the next-to-leading order 
corrections are thus corrected by factors of 2 and larger at large $x$ to lower values. Soft resummation
beyond NLO was considered in \cite{SIMULA}.
In the present analysis we limited the 
investigation to the inclusion of the large $x$ terms in N$^4$LO which are still in the 
vicinity of a nearly
complete QCD analysis as outlined above. The present description is likely to be final 
for values of $x \leq 0.8$.
Beyond this range there are only few data. More data in this interesting region would be 
welcome and can be obtained
at planned high-luminosity colliders such as EIC \cite{EIC}. 

\begin{minipage}[h]{0.40\linewidth}
\centering\epsfig{figure=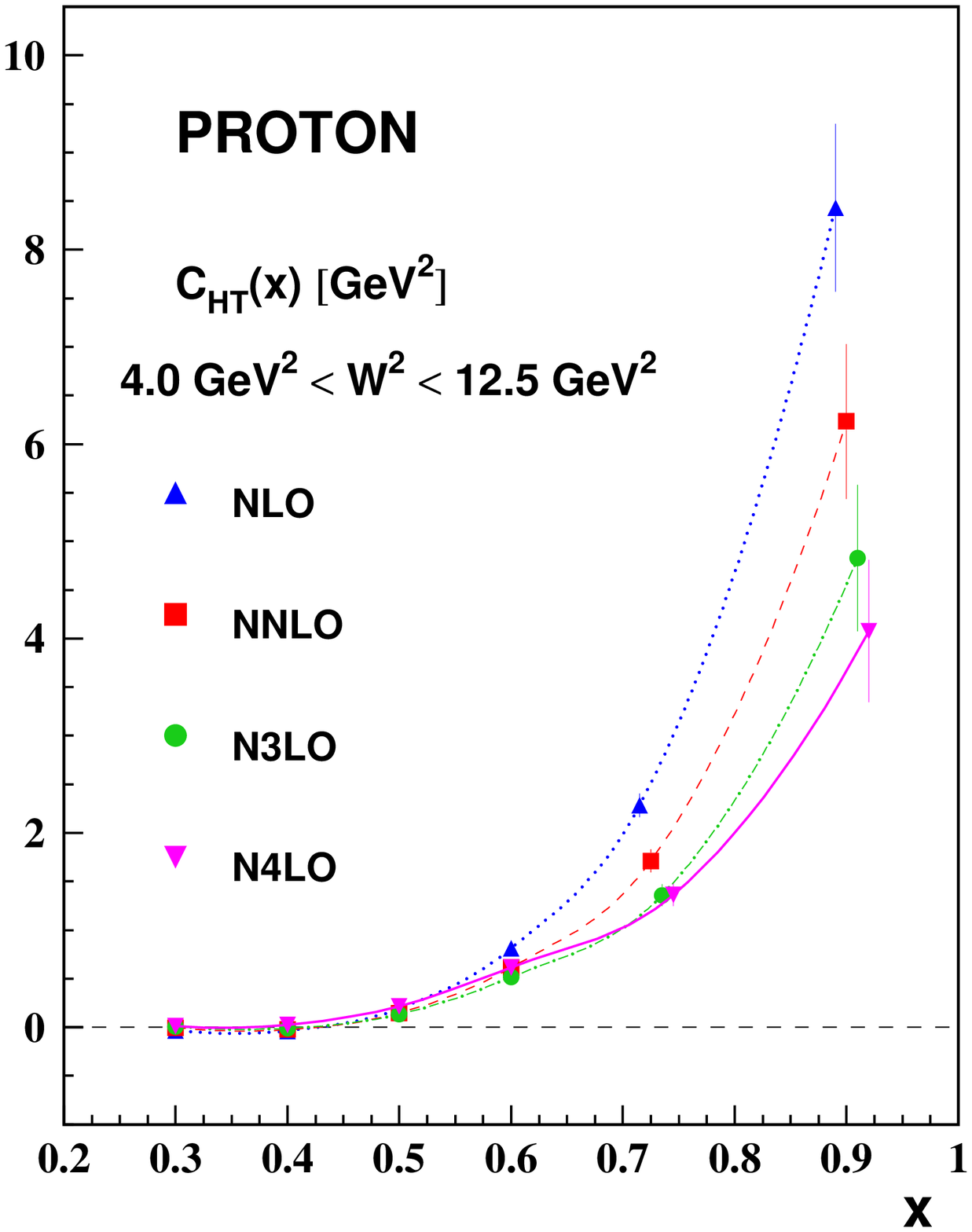,width=\linewidth}
\end{minipage}
\hfill
\begin{minipage}[h]{0.40\linewidth}
\centering\epsfig{figure=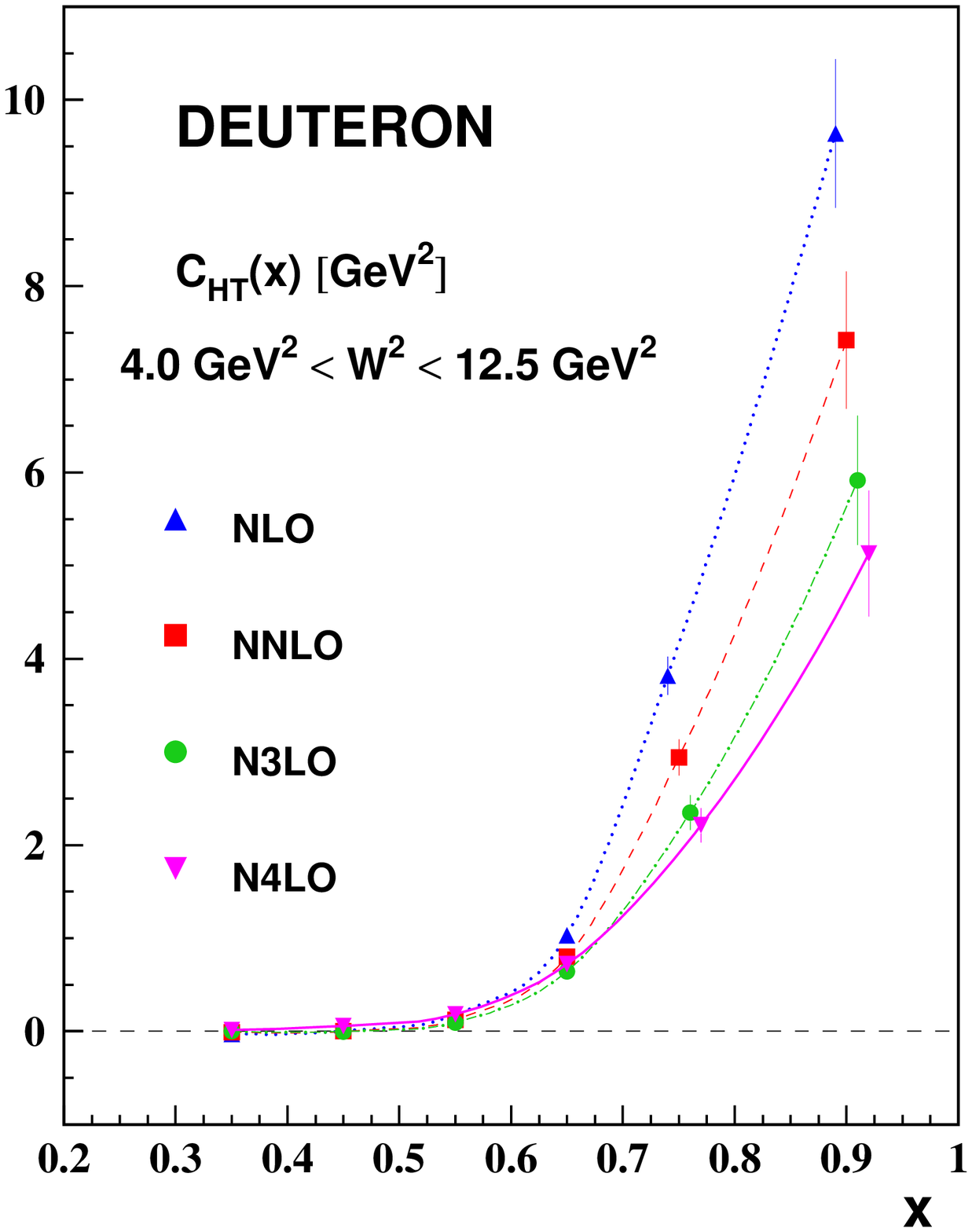,width=\linewidth}
\end{minipage}

\vspace*{2mm}\noindent
{\sf Figure~1~: The coefficient $C_{\rm HT}(x)$ for proton and deuteron data in the large
The curves correspond to the cases of  twist--2 corrections in NLO: 
dotted line, NNLO: dashed line, N$^3$LO dash-dotted line, and asymptotic N$^4$LO: full 
line, cf.~\cite{BBHT}.}

\vspace*{2mm}\noindent
In the present analysis we extracted the large $x$ dynamical higher twist contributions to the structure functions 
in a  model-independent way. It would be interesting to compare moments of the term 
$C_{\rm HT}(x)$ to lattice
results, which allow to simulate the moments of the corresponding higher twist correlation functions, in the future.

%%%%%%%%%%%%%%%%%%%%%%%%%%%%%%%%%%%%%%%%%%%%%%%%%%%%%%%%%%%%%%%%%%%%%%%%%%%%%%%%%%%%%%%%%%
%\section{Bibliography}
\begin{footnotesize}

\end{footnotesize}
\end{document}